\documentclass[conference]{IEEEtran}
\IEEEoverridecommandlockouts
\usepackage{cite}
\usepackage{amsmath,amssymb,amsfonts}
\usepackage{algorithmic}
\usepackage{graphicx}
\usepackage{textcomp}
\usepackage{xcolor}
\usepackage{booktabs} 

\usepackage{multirow}
\usepackage{bm}
\usepackage{amsmath, amscd, amsthm, amssymb, mathrsfs,amsfonts}
\usepackage{graphicx}
\usepackage{color}
\usepackage{mathtools}
\usepackage{subcaption}
\usepackage{tabularx}
\usepackage{array}

\usepackage{microtype}
\usepackage[american]{babel}

\def\b{{\bf b}}

\def\x{{\bf x}}

\def\E{{\bf E}}
\def\F{{\bf F}}

\def\K{{\bf K}}

\def\M{{\bf M}}
\def\N{{\bf N}}

\def\P{{\bf P}}
\def\Q{{\bf Q}}

\def\S{{\bf S}}

\def\U{{\bf U}}
\def\V{{\bf V}}
\def\W{{\bf W}}

\def\0{{\bf 0}}
\def\1{{\bf 1}}
\def\2{{\bf 2}}
\def\3{{\bf 3}}
\def\4{{\bf 4}}
\def\5{{\bf 5}}
\def\6{{\bf 6}}
\def\7{{\bf 7}}
\def\8{{\bf 8}}
\def\9{{\bf 9}}

\newcolumntype{x}[1]{>{\centering\arraybackslash\hspace{0pt}}p{#1}}

\def\BibTeX{{\rm B\kern-.05em{\sc i\kern-.025em b}\kern-.08em
    T\kern-.1667em\lower.7ex\hbox{E}\kern-.125emX}}
    
\newcommand{\xhdr}[1]{\subsubsection*{\bf #1}}
\begin{document}

\title{Self-Attentive Sequential Recommendation}

\author{\IEEEauthorblockN{Wang-Cheng Kang, Julian McAuley}
UC San Diego\\
\{wckang,jmcauley\}@ucsd.edu}

\maketitle

\begin{abstract}
Sequential dynamics are a key feature of many modern recommender systems, which seek to capture the `context' of users' activities on the basis of actions they have performed recently.
To capture such patterns, 
two approaches have proliferated:
Markov Chains (MCs) and Recurrent Neural Networks (RNNs). Markov Chains assume that 
a user's next action can be predicted on the basis of just their last (or last few) actions, while RNNs in principle allow for longer-term semantics to be uncovered.
Generally speaking, MC-based methods perform best in extremely sparse datasets, where model parsimony is critical, while RNNs perform better in denser datasets where higher model complexity is affordable.
The goal of our work is to balance these two goals, by proposing a self-attention based sequential model (SASRec) that allows us to capture long-term semantics (like an RNN), but, using an attention mechanism, makes its predictions based on relatively few actions (like an MC).
At each time step, SASRec seeks to identify 
which items are `relevant' from a user's action history,
and use them to predict the next item. Extensive empirical studies show that our method 
outperforms various state-of-the-art sequential models (including MC/CNN/RNN-based approaches) on both sparse and dense datasets. Moreover, the model is 
an order of magnitude more
efficient than comparable CNN/RNN-based models.
Visualizations on attention weights also show how our model adaptively handles datasets with various density, and uncovers meaningful patterns in activity sequences.
\end{abstract}



\section{Introduction}


The goal of sequential recommender systems is to combine personalized models of user behavior (based on historical activities) with some notion of `context' on the basis of users' recent actions. Capturing useful patterns from sequential dynamics is challenging, primarily because the dimension of the input space grows exponentially with the number of past actions used as context. Research in sequential recommendation is therefore largely concerned with how to capture these high-order dynamics succinctly.

Markov Chains (MCs) are a classic example, 
which assume that the next action is conditioned on only the previous action (or previous few),
and have been successfully adopted to
characterize short-range item transitions for recommendation~\cite{rendle2010fpmc}. Another line of work uses Recurrent Neural Networks~(RNNs) to summarize all previous 
actions via
a hidden state, 
which is used to
predict the next action~\cite{DBLP:journals/corr/HidasiKBT15}.

Both approaches, while strong in specific cases, are somewhat limited to certain types of data. MC-based methods, by making strong simplifying assumptions, perform well in high-sparsity settings, but may fail to capture the intricate dynamics of more complex scenarios. Conversely RNNs, while expressive, require large amounts of data (an in particular \emph{dense} data) before they can outperform simpler baselines.

Recently, a new sequential model \emph{Transfomer} achieved state-of-the-art performance and efficiency for machine translation tasks~\cite{transform}. Unlike existing sequential models that use convolutional or recurrent modules, Transformer is purely based on a proposed attention mechanism called `self-attention,' which is highly efficient and capable of uncovering 
syntactic and semantic patterns between words in a sentence. 

\begin{figure}[t]
\centering
\includegraphics[width=.9\linewidth]{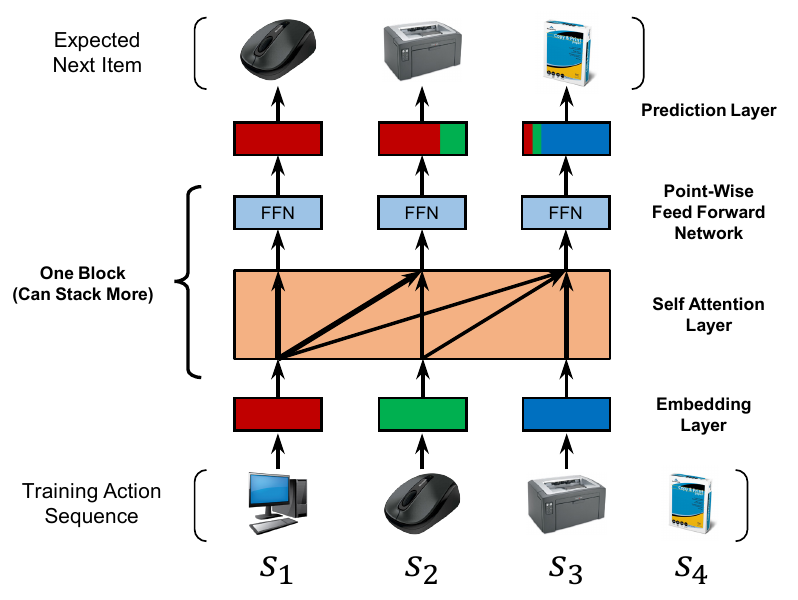}
\caption{A simplified diagram showing the training process of SASRec. At each time step, the model considers all previous items, and 
uses attention to `focus on' items relevant to the
next action.}
\label{fig:diagram}
\end{figure}

Inspired by this method, we seek to apply self-attention mechanisms to sequential recommendation problems.
Our hope is that this idea can address both of the problems outlined above, being on the one hand able to draw context from all actions 
in the past (like RNNs) but on the other hand being able to frame predictions in terms of just a small number of actions (like MCs).
Specifically,
we build a Self-Attention based Sequential Recommendation model (\emph{SASRec}), which adaptively assigns weights to previous items at each time step (Figure \ref{fig:diagram}). 

The proposed model significantly outperforms state-of-the-art MC/CNN/RNN-based sequential recommendation methods on several benchmark datasets. In particular, we examine performance as a function of dataset sparsity, where model performance aligns closely with the patterns described above.
Due to the self-attention mechanism, SASRec tends to consider long-range dependencies on dense datasets, while focusing on more recent activities on sparse datasets. This proves crucial for adaptively handling datasets with varying density.

Furthermore, the core component (i.e., the self-attention block) of SASRec is suitable for parallel acceleration, 
resulting in a model that is an order of magnitude faster
than CNN/RNN-based alternatives.
In addition, we analyze the complexity and scalability of SASRec, conduct a comprehensive ablation study to show the effect of key components, and visualize the attention weights to qualitatively reveal the model's behavior.

\section{Related Work}

Several lines of work 
are closely related to ours.
We first discuss general, followed by temporal, recommendation, before discussing sequential recommendation (in particular MCs and RNNs). Last we
introduce the attention mechanism, especially the self-attention module which is at the core of our model.

\subsection{General Recommendation}

Recommender systems focus on modeling the compatibility between users and items, based on historical feedback (e.g.~clicks, purchases, likes). User feedback can be \emph{explicit} (e.g.~ratings) or \emph{implicit} (e.g.~clicks, purchases, comments)~\cite{WRMF,rendle2009bpr}. Modeling
implicit feedback can be challenging due to the ambiguity of interpreting `non-observed' (e.g.~non-purchased) data.
To address the problem, \emph{point-wise}~\cite{WRMF} and \emph{pairwise}~\cite{rendle2009bpr} methods are proposed to solve such challenges.

Matrix Factorization (MF) methods seek to uncover latent dimensions to represent users' preferences and items' properties,
and estimate interactions through the inner product between the user and item embeddings \cite{Handbook,korenSurvey}.
In addition, another line of work is based on Item Similarity Models~(ISM) and doesn't explicitly model each user with latent factors (e.g.~FISM~\cite{kabbur2013fism}). They learn 
an
item-to-item similarity matrix, and estimate a user's preference toward an item via measuring its similarities with items that the user has interacted with before.

Recently, 
due to their success in related problems,
various deep learning techniques have been introduced for recommendation~\cite{DBLP:journals/corr/ZhangYS17aa}. One line of work seeks to use neural networks to extract item features (e.g.~images~\cite{wang2017your,DBLP:conf/icdm/KangFWM17}, text~\cite{DBLP:conf/kdd/WangWY15,DBLP:conf/recsys/KimPOLY16}, etc.) for content-aware recommendation. Another line of work seeks to replace conventional MF.
For example,
NeuMF~\cite{NeuMF} estimates user preferences via Multi-Layer Perceptions~(MLP), and AutoRec~\cite{sedhain2015autorec} predicts ratings using autoencoders.

\subsection{Temporal Recommendation}	

Dating back to the \emph{Netflix} Prize, temporal recommendation has shown strong performance on various tasks by explicitly modeling the timestamp of users' activities.
TimeSVD++~\cite{timeSVD} achieved 
strong
results by 
splitting
time into several segments and modeling users and items 
separately in each.
Such models are essential
to understand 
datasets that exhibit significant (short- or long-term) temporal `drift'
(e.g.~
`how have movie preferences changed in the last 10 years,' or
`what 
kind of businesses do users visit at 4pm?', etc.)~%
\cite{DBLP:conf/wsdm/WuABSJ17,xiong2010temporal,timeSVD}. Sequential recommendation (or next-item recommendation) differs 
slightly from this setting,
as it only considers the order of actions, and models sequential patterns which are independent of time.
Essentially, sequential models try to model the `context' of users' actions based on their recent activities, rather than considering temporal patterns
\emph{per se}.

\subsection{Sequential Recommendation}

Many sequential recommender systems seek to model 
item-item transition matrices
as a means of capturing sequential patterns among
successive items. 
For instance,
FPMC fuses an MF term and an item-item transition term to capture long-term preferences and short-term transitions respectively~\cite{rendle2010fpmc}. Essentially, the captured transition is a first-order Markov Chain (MC), whereas 
higher-order MCs
assume the next 
action is
related to 
several previous actions.
Since the last visited item is often the key factor affecting the user's next action (essentially providing `context'), first-order MC based methods show strong performance, especially on sparse datasets~\cite{DBLP:conf/recsys/HeKM17}.
There are also methods adopting high-order MCs that consider more previous items~\cite{DBLP:conf/recsys/HeFWM16,DBLP:conf/icdm/HeM16}. In particular, Convolutional Sequence Embedding~(Caser), a 
CNN-based method, views the embedding matrix of $L$ previous items as an `image' and applies convolutional operations to extract transitions~\cite{DBLP:conf/wsdm/TangW18}.

Other than MC-based methods, another line of work 
adopts RNNs to model user sequences~\cite{DBLP:conf/wsdm/JingS17,DBLP:journals/corr/HidasiKBT15,DBLP:conf/icdm/LiuWWLW16,DBLP:conf/wsdm/BeutelCJXLGC18}. For example, GRU4Rec uses Gated Recurrent Units (GRU) to model click sequences for session-based recommendation~\cite{DBLP:journals/corr/HidasiKBT15}, and an improved version further boosts its Top-N recommendation performance~\cite{DBLP:journals/corr/HidasiK17}. In each time step, RNNs take the state from the last step and current action as its input. 
These dependencies
make RNNs less efficient,
though techniques like `session-parallelism' 
have been
proposed to improve efficiency~\cite{DBLP:journals/corr/HidasiKBT15}.

\subsection{Attention Mechanisms}

Attention mechanisms
have
been shown 
to be
effective in various tasks such as image captioning~\cite{DBLP:conf/icml/XuBKCCSZB15} and
machine translation~\cite{DBLP:journals/corr/BahdanauCB14}, among others. 
Essentially the idea behind such mechanisms is that sequential outputs (for example) each depend on `relevant' parts of some input that the model should focus on successively.
An additional benefit is that attention-based methods are 
often
more interpretable. Recently, 
attention mechanisms have been
incorporated into
recommender systems~\cite{DBLP:conf/sigir/ChenZ0NLC17, DBLP:conf/ijcai/XiaoY0ZWC17, DBLP:conf/aaai/WangHCHL018}. For example,
Attentional Factorization Machines (AFM)~\cite{DBLP:conf/ijcai/XiaoY0ZWC17} learn the importance of each feature interaction for content-aware recommendation. 


However, the attention technique used in the above is essentially an \emph{additional} component to the
original
model (e.g.~attention+RNNs, attention+FMs, etc.). Recently, a purely attention-based sequence-to-sequence method, Transfomer~\cite{transform}, achieved state-of-the-art performance and efficiency on machine translation tasks which 
had previously been
dominated by RNN/CNN-based approaches~\cite{wu2016google,DBLP:journals/tacl/ZhouCWLX16}. The Transformer model 
relies
heavily 
on
the proposed `self-attention' modules to capture complex structures 
in
sentences, and to retrieve relevant words (in the source language) for generating the next word (in the target language). Inspired by Transformer, we seek to build a new sequential recommendation model based upon the self-attention approach, 
though the problem of sequential recommendation is quite different from machine translation, 
and 
requires specially designed models.

\section{Methodology}

\begin{table}[t]
\caption{Notation. \label{tb:notation}}
\begin{tabularx}{\linewidth}{lX}
\toprule
Notation&Description\\
\midrule
$\mathcal{U},\mathcal{I}$        & user and item set\\
$\mathcal{S}^u$                 & historical interaction sequence for a user $u$: $(\mathcal{S}^u_1, \mathcal{S}^u_2, ... , \mathcal{S}^u_{|\mathcal{S}^u|})$ \\
$d\in \mathbb{N}$                 & latent vector dimensionality\\
$n\in \mathbb{N}$                 & maximum sequence length\\
$b\in \mathbb{N}$                 & number of self-attention blocks\\
$\M\in\mathbb{R}^{|\mathcal{I}|\times d}$        & item embedding matrix\\
$\P\in\mathbb{R}^{n\times d}$        & positional embedding matrix\\
$\widehat{\E}\in\mathbb{R}^{n\times d}$        & input embedding matrix\\
$\S^{(b)}\in\mathbb{R}^{n\times d}$   & item embeddings after the $b$-th self-attention layer\\
$\F^{(b)}\in\mathbb{R}^{n\times d}$   & item embeddings after the $b$-th feed-forward network\\
\bottomrule
\end{tabularx}
\end{table}

In the setting of sequential recommendation, we are given a user's action sequence $\mathcal{S}^u=(\mathcal{S}^u_1, \mathcal{S}^u_2, \ldots , \mathcal{S}^u_{|\mathcal{S}^u|})$, and seek to predict the next item. During the training process, at time step $t$, the model predicts the next item depending 
on 
the
previous $t$ items. 
As shown in Figure \ref{fig:diagram},
it will be convenient to think of the model's input as
$(\mathcal{S}^u_1, \mathcal{S}^u_2, \ldots , \mathcal{S}^u_{|\mathcal{S}^u|-1})$ 
and its expected output as a `shifted' version of the same sequence:
$(\mathcal{S}^u_2, \mathcal{S}^u_3, \ldots , \mathcal{S}^u_{|\mathcal{S}^u|})$. In this section, we describe 
how we build a sequential recommendation model 
via an
embedding layer, several self-attention blocks, and a prediction layer. We also analyze its complexity and further discuss how SASRec differs from related models.
%
Our notation is summarized in Table \ref{tb:notation}.

\subsection{Embedding Layer}

We transform the training sequence $(\mathcal{S}^u_1, \mathcal{S}^u_2, ... , \mathcal{S}^u_{|\mathcal{S}^u|-1})$ into a fixed-length sequence $s=(s_1,s_2,\dots,s_n)$, where $n$ represents the maximum length that our model can handle. If the sequence length is greater than $n$, we 
consider
the most recent $n$ actions. If the sequence length is less than $n$, we 
repeatedly
add a `padding' item to the left until the length is $n$. We create an item embedding matrix $\M\in\mathbb{R}^{|\mathcal{I}|\times d}$ where $d$ is the latent dimensionality, and retrieve the input embedding matrix $\E\in\mathbb{R}^{n\times d}$, where $\E_i=\M_{s_i}$.
A constant zero vector $\0$ is used as the embedding for the padding item.

\xhdr{Positional Embedding}
As we will see in the next section, since the self-attention model doesn't include any recurrent or convolutional module, it is not aware of the positions of previous items. Hence we inject a learnable position embedding $\P\in\mathbb{R}^{n\times d}$ into the input embedding:
\begin{equation}
\widehat{\E}=\left [
  \begin{tabular}{c}
  $\M_{s_1}+\P_{1}$ \\
  $\M_{s_2}+\P_{2}$ \\
  $\dots$ \\
  $\M_{s_n}+\P_{n}$
  \end{tabular}
\right ]
\end{equation}
We also tried the fixed position embedding as used in~\cite{transform}, 
but found that this led to worse performance in our case.
We analyze the effect of the position embedding quantitatively and qualitatively in our experiments.

\subsection{Self-Attention Block}

The scaled dot-product attention~\cite{transform} is defined as:
\begin{equation}
\text{Attention}(\Q,\K,\V)=\text{softmax}\left (\frac{\Q\K^T}{\sqrt{d}}\right )\V,
\end{equation}
where $\Q$ represents the queries, $\K$ 
the keys and $\V$ 
the values (each row represents an item). Intuitively, the attention layer calculates a weighted sum of all values, where the weight between query $i$ and value $j$ 
relates to
the interaction between query $i$ and key $j$. The scale factor $\sqrt{d}$ is to avoid 
overly large
values of the inner product, especially when the dimensionality is high.

\xhdr{Self-Attention layer} In NLP tasks 
such as
machine translation,
attention mechanisms
are typically used
with $\K=\V$ (e.g.~using an RNN encoder-decoder for translation: the encoder's
hidden states
are keys and values, and the decoder's
hidden states
are queries)~\cite{DBLP:journals/corr/BahdanauCB14}. 
Recently, a self-attention method was proposed which uses the same objects as queries, keys, and values~\cite{transform}. In our case, the self-attention operation takes the embedding $\widehat{\E}$ as input, converts it to three matrices through linear projections, and feeds them into an attention layer:
\begin{equation}
\S=\text{SA}(\widehat{\E})=\text{Attention}(\widehat{\E}\W^Q,\widehat{\E}\W^K,\widehat{\E}\W^V),
\label{eq:sa}
\end{equation}
where the projection matrices $\W^Q,\W^K,\W^V\in\mathbb{R}^{d\times d}$. The projections make the model more flexible. For example, the model can learn asymmetric interactions (i.e.,~\textless query $i$, key $j$\textgreater~and \textless query $j$, key $i$\textgreater~can have different interactions).

\xhdr{Causality} Due to the nature of sequences, the model should consider only the first $t$ items when
predicting
the $(t+1)$-st item. However, the $t$-th output of the self-attention layer ($\S_t$) contains embeddings of subsequent items, which makes the model 
ill-posed.
Hence, we modify the attention by forbidding all links between $\Q_i$ and $\K_j$ ($j>i$).

\xhdr{Point-Wise Feed-Forward Network} Though the self-attention is able to aggregate all previous items' embeddings with adaptive weights, ultimately 
it is
still a linear model. To endow the model with 
nonlinearity and 
to
consider interactions between different latent dimensions, 
we apply
a point-wise two-layer feed-forward network 
to all $\S_i$
identically (sharing parameters):
\begin{equation}
\F_i = \text{FFN}(\S_i)=\text{ReLU}(\S_i\W^{(1)}+\b^{(1)})\W^{(2)}+\b^{(2)},
\label{eq:ffn}
\end{equation}
where $\W^{(1)},\W^{(2)}$ are $d\times d$ matrices and $\b^{(1)},\b^{(2)}$ are $d$-dimensional vectors. Note that 
there is no interaction between $\S_i$ and $\S_j$ ($i\neq j$), meaning that we still prevent 
information leaks (from back to front).

\subsection{Stacking Self-Attention Blocks}

After the first self-attention block, $\F_i$ essentially aggregates all previous items' embeddings (i.e., $\widehat{\E}_j, j\leq i$). However, it might be useful to learn more complex item transitions via 
another self-attention block based on $\F$. Specifically, we stack the self-attention block (i.e., a self-attention layer and a feed-forward network), and the $b$-th ($b>1$) block is defined as:
\begin{equation}
\begin{array}{c}
\S^{(b)}=\text{SA}(\F^{(b-1)}),\\
\F_i^{(b)}=\text{FFN}(\S_i^{(b)}),\ \ \forall i\in \{1,2,\dots,n\},
\end{array}
\end{equation}
and the $1$-st block is defined as $\S^{(1)}=\S$ and $\F^{(1)}=\F$.

However, when the network goes deeper, several problems become 
exacerbated:
1) the 
increased
model capacity leads to 
overfitting; 2) the training process 
becomes
unstable (due to vanishing gradients etc.); and
3) models with more parameters often require more training time. Inspired by \cite{transform}, We perform the following operations to alleviate these problems:
\[g(x)=x+\text{Dropout}(g(\text{LayerNorm}(x))),\]
where $g(x)$ represents the self attention layer or the feed-forward network. That is to say, for layer $g$ in each block, we apply 
layer normalization on the input $x$ before feeding into $g$, apply dropout on $g$'s output, and add the input $x$ to the final output. We introduce these operations
below.

\xhdr{Residual Connections} In some cases, multi-layer neural networks have 
demonstrated
the ability to learn meaningful features hierarchically~\cite{zeiler2014visualizing}. However, 
simply
adding more layers 
did not easily correspond to
better performance until residual networks
were
proposed~\cite{he2016deep}. 
The core idea behind residual networks is to
propagate
low-layer features to 
higher layers by 
residual 
connection. Hence, if low-layer features are useful, the model can easily propagate 
them
to the final layer. Similarly, we assume 
residual connections are also useful in our case. For example, existing sequential recommendation methods have shown that the last visited item plays a key role on predicting the next item~\cite{rendle2010fpmc,DBLP:conf/icdm/HeM16,DBLP:conf/recsys/HeKM17}. However, after several self-attention blocks, the embedding of the last visited item is entangled with all previous items;
adding residual connections to propagate the last visited item's embedding to the final layer would make it much easier for the model to leverage low-layer information.

\xhdr{Layer Normalization} Layer normalization is 
used
to normalize the inputs across features (i.e., zero-mean and unit-variance), which is beneficial for stabilizing and accelerating 
neural network training~\cite{DBLP:journals/corr/BaKH16}. Unlike 
batch normalization~\cite{DBLP:conf/icml/IoffeS15}, the statistics used in layer normalization are independent of other samples in the same batch. Specifically, assuming the input is a vector $\x$ which contains all features of a sample, the operation is defined as:
\[\text{LayerNorm}(\x)=\bm{\alpha}\odot\frac{\x-\mu}{\sqrt{\sigma^2+\epsilon}}+\bm{\beta},\]
where $\odot$ is an element-wise product (i.e., the Hadamard product), $\mu$ and $\sigma$ 
are
the mean and variance of $\x$, $\bm{\alpha}$ and $\bm{\beta}$ are learned scaling factors and bias terms.

\xhdr{Dropout} To alleviate 
overfitting problems in deep neural networks, 
`Dropout' regularization techniques have
been shown to be effective in various neural network
architectures~\cite{DBLP:journals/jmlr/SrivastavaHKSS14}. The idea of dropout is simple: randomly
`turn off'
neurons with probability $p$ during
training, and 
use 
all neurons when testing. Further analysis points out that dropout can be viewed as a form of ensemble learning which considers
an
enormous number of models (exponential in the number of neurons and input features) that share parameters~\cite{DBLP:journals/corr/Warde-FarleyGCB13}. We also apply a dropout layer on the embedding $\widehat{\E}$.

\subsection{Prediction Layer}
After $b$ self-attention blocks that adaptively and hierarchically extract information of previously 
consumed
items, we predict the next item (given the first $t$ items) based on $\F_t^{(b)}$. Specifically, we adopt an MF layer to predict the relevance of item $i$:
\[r_{i,t}=\F_t^{(b)}\N_i^T,\]
where $r_{i,t}$ is the relevance of item $i$ being the next item given the first $t$ items (i.e., $s_1,s_2,\ldots,s_t$), and $\N\in\mathbb{R}^{|\mathcal{I}|\times d}$ is an item embedding matrix. Hence, a high interaction score $r_{i,t}$ means a high relevance, and we can generate recommendations by ranking the scores.

\xhdr{Shared Item Embedding} To reduce the model size and alleviate overfitting, we consider another scheme which only uses a single item embedding $\M$:
\begin{equation}
r_{i,t}=\F_t^{(b)}\M_i^T.
\end{equation}
Note that $\F_t^{(b)}$ can be represented as a function depending on 
the
item embedding $\M$: $\F_t^{(b)}=f(\M_{s_1},\M_{s_2},\dots,\M_{s_t})$. A potential issue of using homogeneous item embeddings is that their inner products cannot represent asymmetric item transitions (e.g.~item $i$ is frequently bought after $j$, but not vise versa), and thus existing methods like FPMC tend to use heterogeneous item embeddings. However, our model doesn't have this issue since it learns a nonlinear transformation. For example, the feed forward network can easily achieve the asymmetry with the same item embedding: $\text{FFN}(\M_i)\M_j^T\neq \text{FFN}(\M_j)\M_i^T$. Empirically, 
using a shared item embedding 
significantly improves the performance of our model.

\xhdr{Explicit User Modeling} To provide personalized recommendations, existing methods often take one of two approaches: 1) learn an \emph{explicit} user embedding representing user preferences~(e.g. MF~\cite{MF}, FPMC~\cite{rendle2010fpmc} and Caser~\cite{DBLP:conf/wsdm/TangW18}); 2)~consider the user's previous actions, 
and induce an \emph{implicit} user embedding from embeddings of visited items (e.g.~FSIM~\cite{kabbur2013fism}, Fossil~\cite{DBLP:conf/icdm/HeM16}, GRU4Rec~\cite{DBLP:journals/corr/HidasiKBT15}). Our method falls into the latter category, as we generate an embedding $\F^{(b)}_n$ by considering all actions of a user. However, we can also insert an explicit user embedding at the last layer, for example via addition: $r_{u,i,t}=(\U_u+\F_t^{(b)})\M_i^T$ where $\U$ is 
a
user embedding matrix. However, we empirically find 
that
adding 
an
explicit user embedding doesn't improve 
performance (presumably because the model already considers
all of
the user's
actions). 

\subsection{Network Training}

Recall that we convert each user sequence (excluding the last action) $(\mathcal{S}^u_1, \mathcal{S}^u_2, \ldots , \mathcal{S}^u_{|\mathcal{S}^u|-1})$ to a fixed length sequence $s=\{s_1,s_2,\dots,s_n\}$ 
via truncation
or padding items. We define $o_t$ as the expected output at time step $t$: 
\[o_t=\begin{cases}
\texttt{<pad>}									&	\text{if } s_t \text{ is a padding item}\\
s_{t+1}									&	1\leq t<n\\
\mathcal{S}^{u}_{|\mathcal{S}^u|}		&	t=n
\end{cases},\]
where $\texttt{<pad>}$ indicates a padding item. Our model takes a sequence $s$ as input, the corresponding sequence $o$ as expected output, and we adopt the binary cross entropy loss as the objective function:
\[-\sum_{\mathcal{S}^{u}\in\mathcal{S}}\sum_{t\in[1,2,\dots,n]}\left [\log(\sigma(r_{o_t,t}))+\sum_{j\not\in \mathcal{S}^{u}}\log(1-\sigma(r_{j,t}))\right ].\]
Note that we ignore the terms where $o_t=\texttt{<pad>}$.

The network is optimized by the \emph{Adam} optimizer~\cite{DBLP:journals/corr/KingmaB14}, which is a variant of Stochastic Gradient Descent (SGD) with adaptive moment estimation. In each epoch, we randomly generate one negative item $j$ for each time step in each sequence. 
More implementation details 
are described later.

\subsection{Complexity Analysis}
\label{sec:complexity}

\xhdr{Space Complexity} The learned parameters in our model are from the embeddings and parameters in the self-attention layers, feed-forward networks and layer normalization. The total number of parameters is $O(|\mathcal{I}|d+nd+d^2)$, which is moderate compared to other methods (e.g.~$O(|\mathcal{U}|d+|\mathcal{I}|d)$ for FPMC) since 
it does not grow with
the number of users, and $d$ is typically small in recommendation problems.

\xhdr{Time Complexity} The computational complexity of our model is mainly 
due to
the self-attention layer and the feed-forward network, which is $O(n^2d+nd^2)$. The dominant term is typically $O(n^2d)$ from the self-attention layer. However, a 
convenient
property in our model is that the computation in each self-attention layer is fully parallelizable, which is 
amenable to
GPU acceleration. In contrast, RNN-based methods (e.g.~GRU4Rec~\cite{DBLP:journals/corr/HidasiKBT15}) have a
dependency on time steps (i.e., computation on time step $t$ must wait for results from time step $t\text{-}1$), which leads to an $O(n)$ time on sequential operations. 
We empirically find our method is over ten times faster than RNN and CNN-based methods with GPUs (the result is similar to that in \cite{transform} for machine translation tasks), and the maximum length $n$ can easily scale to a few hundred which is generally sufficient for existing benchmark datasets.

During testing, for each user, after calculating the embedding $\F_n^{(b)}$
, the 
process is the same as standard MF methods. ($O(d)$ for evaluating the preference toward an item).

\xhdr{Handing 
Long Sequences} Though our experiments empirically verify the efficiency of our method, ultimately it cannot scale to very long sequences. A few options are promising to 
investigate
in the future: 1) using 
restricted self-attention~\cite{poveytime} which only attends on
recent actions rather than all actions, and distant actions can be considered in higher layers; 2) splitting long sequences into short segments as in \cite{DBLP:conf/wsdm/TangW18}.

\subsection{Discussion}

We find
that
SASRec can be viewed as a generalization 
of some classic CF models. We also discuss 
conceptually
how our approach and existing methods handle sequence modeling.

\xhdr{Reduction to Existing Models}

\begin{itemize}
\item \emph{Factorized Markov Chains}: FMC factorizes a first-order item transition matrix, and predicts the next item $j$ depending on the last item $i$:
\[P(j|i)\propto\M_i^T\N_j.\]
If we set 
the
self-attention block to zero, use unshared item embeddings, and remove the position embedding, SASRec 
reduces
to FMC.
Furthermore, SASRec is also closely related to Factorized Personalized Markov Chains~(FPMC)~\cite{rendle2010fpmc}, which fuse MF with FMC to capture user preferences and short-term dynamics respectively:
\[P(j|u,i)\propto\left [\U_u,\M_i\right ]\N^T_j.\]
Following the reduction operations above for FMC, and adding an explicit user embedding (via concatenation), SASRec is equivalent to FPMC.


\item \emph{Factorized Item Similarity Models~\cite{kabbur2013fism}}: FISM estimates a preference score 
toward
item $i$ by considering the similarities between $i$ and items the user consumed before:
\[P(j|u)\propto\left (\frac{1}{|\mathcal{S}^u|}\sum_{i\in\mathcal{S}^u}\M_i\right)\N^T_j.\]

If we use 
one
self-attention layer (excluding the feed-forward network), set uniform attention weights (i.e., $\frac{1}{|\mathcal{S}_{u}|}$) on items, use unshared item embeddings, and remove the position embedding, SASRec is reduced to FISM. Thus our model can be viewed as an adaptive, hierarchical, sequential item similarity model for next item recommendation. 

\end{itemize}

\xhdr{MC-based Recommendation} Markov Chains (MC) can effectively capture local sequential patterns, 
assuming  that
the next item is only 
dependent
on the previous $L$ items. Exiting MC-based sequential recommendation methods rely on either first-order MCs (e.g.~FPMC~\cite{rendle2010fpmc}, HRM~\cite{hrm}, TransRec~\cite{DBLP:conf/recsys/HeKM17}) or high-order MCs (e.g.~Fossil~\cite{DBLP:conf/icdm/HeM16}, Vista~\cite{DBLP:conf/recsys/HeFWM16}, Caser~\cite{DBLP:conf/wsdm/TangW18}). 
The first group of methods
tend to perform best on sparse datasets.
In contrast,
higher-order MC based methods
have two limitations: (1) The MC order $L$ needs to be specified before 
training, 
rather than being chosen
adaptively; (2) The performance and efficiency doesn't scale well 
with the
order $L$, 
hence $L$ is typically small
(e.g.~less than 5). 
Our method 
resolves
the first issue, since it can adaptively attend on related previous items (e.g.~focusing on just the last item 
on sparse dataset, and 
more items on dense datasets). Moreover, our model is essentially conditioned on $n$ previous items, and can empirically scale to 
several
hundred previous items,
exhibiting performance gains with moderate increases in training time.

\xhdr{RNN-based Recommendation} Another line of work seeks to use RNNs to model user action sequences~\cite{DBLP:journals/corr/HidasiKBT15,DBLP:journals/corr/HidasiK17,DBLP:conf/wsdm/WuABSJ17}. RNNs are generally suitable for modeling sequences, though recent studies show CNNs and self-attention can be stronger in some sequential settings~\cite{transform,DBLP:journals/corr/abs-1803-01271}. Our self-attention based model can be derived from item similarity models, which are a reasonable alternative for sequence modeling for recommendation. For RNNs, other than their inefficiency in parallel computation (%
Section \ref{sec:complexity}), 
their
maximum path length (from an input node to related output nodes) is $O(n)$. In contrast, our model has $O(1)$ maximum path length, 
which can be beneficial
for learning long-range dependencies~\cite{hochreiter2001gradient}. 

\section{Experiments}

In this section, we present 
our
experimental setup and empirical results. Our experiments are designed to answer the following research questions:
\begin{description}
\item[\textbf{RQ1:}] Does 
SASRec
outperform state-of-the-art models including CNN/RNN based methods?
\item[\textbf{RQ2:}] What 
is the influence of various
components in the SASRec architecture?
\item[\textbf{RQ3:}] What is the training efficiency and scalability (regarding $n$) of SASRec?
\item[\textbf{RQ4:}] Are the attention weights able to learn meaningful patterns related to positions or items' attributes?
\end{description}

\subsection{Datasets}

We evaluate our methods on four datasets from three real world applications. The datasets 
vary
significantly in domains, platforms, and sparsity: 
\begin{itemize}
\item \textbf{Amazon:} A series of datasets introduced in~\cite{VisualSIGIR}, comprising large corpora of product reviews crawled from \emph{Amazon.com}. Top-level product categories on \emph{Amazon} are treated as separate datasets. We consider two categories, `Beauty,' and `Games.' This dataset is notable for its high sparsity and variability.
\item \textbf{Steam:} We introduce a new dataset crawled from \emph{Steam}, a large online video game distribution platform. The dataset contains 2,567,538 users, 15,474 games and 7,793,069 English reviews spanning October 2010 to January 2018. The dataset also includes rich information that might be useful in future work, like 
users'
play hours, pricing information, media score, category, developer (etc.).
\item \textbf{MovieLens:} A widely used benchmark dataset for evaluating collaborative filtering algorithms. We use the version (MovieLens-1M) that includes 1 million user ratings.
\end{itemize}

We followed the same preprocessing procedure from \cite{DBLP:conf/recsys/HeKM17,DBLP:conf/icdm/HeM16,rendle2010fpmc}. For all datasets, we treat the presence of a review or rating as implicit feedback (i.e., the user interacted with the item) and use timestamps to determine the sequence order of actions. We discard users and items with fewer than 5 related actions. For partitioning, we split the historical sequence $\mathcal{S}^u$ for each user $u$ into three parts: (1) the most recent action $\mathcal{S}^u_{|\mathcal{S}^u|}$ for testing, (2) the second most recent action $\mathcal{S}^u_{|\mathcal{S}^u|-1}$ for validation, and (3) all remaining actions for training. Note that during testing, the input sequences contain training actions and the validation action.

Data statistics are shown in Table \ref{tb:data}.
We 
see that the two \emph{Amazon} datasets have the 
fewest
actions per user and per item (on average), \emph{Steam} has a high average number of actions per item, and \emph{MovieLens-1m} is the most dense dataset.

\begin{table}[h]
\centering
\caption{Dataset statistics (after preprocessing) \label{exp:dataset}}
\label{tb:data}
\setlength{\tabcolsep}{5pt}
\begin{tabular}{lrrrrr}
\toprule
\multicolumn{1}{x{0.8cm}}{\newline Dataset} 				& \multicolumn{1}{x{0.8cm}}{\newline \#users} & \multicolumn{1}{x{0.8cm}}{\newline \#items} & \multicolumn{1}{x{0.8cm}}{avg. actions /user}&\multicolumn{1}{x{0.8cm}}{avg. actions /item} & \multicolumn{1}{x{0.8cm}}{\newline \#actions}\\ \midrule
\emph{Amazon Beauty} 	&  52,024  	&  57,289  & 7.6 	& 6.9	&   0.4M\\
\emph{Amazon Games}   	&  31,013  	&  23,715  & 9.3 	& 12.1  &   0.3M\\
\emph{Steam}     		&  334,730  &  13,047  & 11.0 	& 282.5	&	3.7M\\
\emph{MovieLens-1M}    	&  6,040 	&   3,416  & 163.5 & 289.1	&	1.0M\\ \bottomrule
\end{tabular}
\end{table}

\begin{table*}
\centering
\caption{Recommendation performance. The best performing method in each row is boldfaced, and the second best method in each row is underlined. Improvements over non-neural and neural approaches are shown in the last two columns respectively.
\label{tab:recommendation}}
\begin{tabular}{llccccccccccc}
\toprule
\multirow{2}{*}{Dataset} & \multirow{2}{*}{Metric}     & \multirow{2}{*}{\begin{tabular}[c]{@{}c@{}}(a)\\ PopRec\end{tabular}} & \multirow{2}{*}{\begin{tabular}[c]{@{}c@{}}(b)\\ BPR\end{tabular}} & \multirow{2}{*}{\begin{tabular}[c]{@{}c@{}}(c)\\  FMC\end{tabular}} & \multirow{2}{*}{\begin{tabular}[c]{@{}c@{}}(d)\\ FPMC\end{tabular}} & \multirow{2}{*}{\begin{tabular}[c]{@{}c@{}}(e)\\ TransRec\end{tabular}} & \multirow{2}{*}{\begin{tabular}[c]{@{}c@{}}(f)\\ GRU4Rec\end{tabular}} &\multirow{2}{*}{\begin{tabular}[c]{@{}c@{}}(g)\\ GRU4Rec$^\texttt{+}$ \end{tabular}}& \multirow{2}{*}{\begin{tabular}[c]{@{}c@{}}(h)\\ Caser\end{tabular}} & \multirow{2}{*}{\begin{tabular}[c]{@{}c@{}}(i)\\ SASRec\end{tabular}} &\multicolumn{2}{c}{Improvement vs.}\\
 & & & & & & & & & &&\multicolumn{1}{c}{(a)-(e)} & \multicolumn{1}{c}{(f)-(h)}  \\ \midrule
\multirow{2}{*}{\emph{Beauty}} 			& Hit@10  & 0.4003 & 0.3775 & 0.3771 & 0.4310 & \underline{0.4607} & 0.2125 & 0.3949 & 0.4264 & \textbf{0.4854} & 5.4\%	& 13.8\% \\
                                       	& NDCG@10 & 0.2277 & 0.2183 & 0.2477 & 0.2891 & \underline{0.3020} & 0.1203 & 0.2556 & 0.2547 & \textbf{0.3219} & 6.6\% & 25.9\% \\[1.5mm]
\multirow{2}{*}{\emph{Games}}     		& Hit@10  & 0.4724 & 0.4853 & 0.6358 & 0.6802 & \underline{0.6838} & 0.2938 & 0.6599 & 0.5282 & \textbf{0.7410} & 8.5\%	& 12.3\%  \\
                                        & NDCG@10 & 0.2779 & 0.2875 & 0.4456 & 0.4680 & 0.4557 & 0.1837 & \underline{0.4759} & 0.3214 & \textbf{0.5360} & 14.5\% & 12.6\% \\[1.5mm]
\multirow{2}{*}{\emph{Steam}}      		& Hit@10  & 0.7172 & 0.7061 & 0.7731 & 0.7710 & 0.7624 & 0.4190 & \underline{0.8018} & 0.7874 & \textbf{0.8729} & 13.2\% & 8.9\%\\
                                        & NDCG@10 & 0.4535 & 0.4436 & 0.5193 & 0.5011 & 0.4852 & 0.2691 & \underline{0.5595} & 0.5381 & \textbf{0.6306} & 21.4\% & 12.7\% \\[1.5mm]
\multirow{2}{*}{\emph{ML-1M}}    		& Hit@10  & 0.4329 & 0.5781 & 0.6986 & 0.7599 & 0.6413 & 0.5581 & 0.7501 & \underline{0.7886} & \textbf{0.8245} & 8.5\% & 4.6\%\\
                                        & NDCG@10 & 0.2377 & 0.3287 & 0.4676 & 0.5176 & 0.3969 & 0.3381 & 0.5513 & \underline{0.5538} & \textbf{0.5905} & 14.1\%& 6.6\% \\
\bottomrule
\end{tabular}
\end{table*}

\subsection{Comparison Methods}

To show the effectiveness of our method, we include three groups of recommendation 
baselines.
The first group includes general recommendation methods which only consider user feedback without considering the sequence order of actions:
\begin{itemize}
\item \textbf{PopRec}: This is a simple baseline that ranks items according to their popularity (i.e., number of associated actions).

\item \textbf{Bayesian Personalized Ranking (BPR)~\cite{rendle2009bpr}}: BPR is a classic method 
for
learning personalized rankings from implicit feedback. Biased matrix factorization is used as the underlying recommender.

\end{itemize}

The second group contains sequential recommendation methods based on first order Markov chains, which consider the last visited item:
\begin{itemize}

\item\textbf{Factorized Markov Chains (FMC):} A first-order Markov chain method. FMC factorizes an item transition matrix using two item embeddings, and generates recommendations depending only on the last visited item.

\item\textbf{Factorized Personalized Markov Chains (FPMC)~\cite{rendle2010fpmc}}: FPMC uses a  combination of matrix factorization and factorized first-order Markov chains as its recommender, which captures users' long-term preferences as well as item-to-item transitions.

\item\textbf{Translation-based Recommendation (TransRec)~\cite{DBLP:conf/recsys/HeKM17}}: A state-of-the-art first-order sequential recommendation method which models each user as a translation vector to capture the transition from the current item to the next.
\end{itemize}

The last group contains deep-learning based sequential 
recommender systems,
which consider several (or all) previously visited items:

\begin{itemize}
\item\textbf{GRU4Rec~\cite{DBLP:journals/corr/HidasiKBT15}}: A seminal method 
that uses
RNNs to model user action sequences for session-based recommendation. We treat each user's feedback sequence as a session.

\item\textbf{GRU4Rec$^\text{+}$~\cite{DBLP:journals/corr/HidasiK17}}: An improved version of GRU4Rec, which adopts a different loss function and sampling strategy, and shows significant performance gains on Top-N recommendation.

\item\textbf{Convolutional Sequence Embeddings (Caser)~\cite{DBLP:conf/wsdm/TangW18}}: A recently proposed CNN-based method capturing high-order Markov chains by applying 
convolutional operations on the embedding matrix of the $L$ most recent items, and achieves state-of-the-art sequential recommendation performance.
\end{itemize}

Since other sequential recommendation methods (e.g.~PRME~\cite{feng2015prme}, HRM~\cite{hrm}, Fossil~\cite{DBLP:conf/icdm/HeM16}) have been outperformed on similar datasets by 
baselines among those above,
we omit 
comparison against them. We also don't include temporal recommendation methods like TimeSVD++~\cite{timeSVD} and RRN~\cite{DBLP:conf/wsdm/WuABSJ17}, 
which differ in setting from what we consider here.

For fair comparison, we implement BPR, FMC, FPMC, and TransRec using \emph{TemsorFlow} with the Adam~\cite{DBLP:journals/corr/KingmaB14} optimizer. For GRU4Rec, GRU4Rec$^\text{+}$, and Caser, we use 
code provided by the
corresponding authors. For all methods except PopRec, 
we consider latent dimensions $d$
from $\{10,20,30,40,50\}$. For BPR, FMC, FPMC, and TransRec, the $\ell_2$ regularizer is 
chosen
from $\{0.0001,0.001,0.01,0.1,1\}$. All other hyperparameters and initialization strategies are 
those suggested by the methods' authors.
We tune hyper-parameters using the validation set, and terminate 
training if 
validation performance doesn't improve for 20 epochs.

\subsection{implementation Details}

For the architecture in the default version of SASRec,
we 
use
two
self-attention blocks ($b=2$), and use the learned positional embedding. Item embeddings in the embedding layer and prediction layer are shared.
We implement SASRec with \emph{TensorFlow}. 
The optimizer is the \emph{Adam} optimizer~\cite{DBLP:journals/corr/KingmaB14}, the learning rate is set to $0.001$, and the batch size is $128$. The dropout rate of turning off neurons is 0.2 for \emph{MovieLens-1m} and 0.5 for the other three datasets due to their sparsity. The maximum sequence length $n$ is set to 200 for \emph{MovieLens-1m} and 50 for the other three datasets,
i.e., roughly proportional to the mean number of actions per user.
Performance of variants and different hyper-parameters is examined below.

\emph{All code and data shall be released at publication time}.

\subsection{Evaluation Metrics}

We adopt two common Top-N metrics, Hit Rate@10 and NDCG@10, to evaluate recommendation performance~\cite{NeuMF,DBLP:conf/recsys/HeKM17}.
Hit@10 counts the fraction of times that the ground-truth next item is 
among the
top 10 items, while NDCG@10 is a position-aware metric which assigns larger weights on higher positions. Note that since we only have one test item for each user, Hit@10 is equivalent to Recall@10, and is proportional to Precision@10.

To avoid heavy computation on all user-item pairs, we followed the strategy in \cite{koren2008factorization,NeuMF}. For each user $u$, we randomly sample 100 negative items, and rank these items with the ground-truth item. Based on the rankings of these 101 items, Hit@10 and NDCG@10 can be evaluated.

\subsection{Recommendation Performance}

Table \ref{tab:recommendation} presents the recommendation performance of all methods on the four datasets (\textbf{RQ1}). By considering the second best methods across all datasets, a general pattern emerges with non-neural methods (i.e., (a)-(e)) performing better on sparse datasets and neural approaches (i.e., (f)-(h)) 
performing better on denser datasets.
Presumably 
this owes to
neural approaches having more parameters to capture high order transitions (i.e., they are expressive but easily overfit),
whereas carefully designed but simpler models are more effective in high-sparsity settings.

Our method SASRec outperforms all baselines on both sparse and dense datasets, and gains 6.9\% Hit Rate and 9.6\% NDCG improvements (on average) against the strongest baseline. 
One likely
reason is that our model can adaptively attend items within different ranges on different datasets (e.g.~only 
the previous
item on sparse datasets 
and more
on dense datasets). 
We further analyze this
behavior
in Section \ref{sec:vis}.

\begin{figure*}
\centering
\begin{subfigure}[b]{\textwidth}
\centering
\includegraphics[width=.99\linewidth]{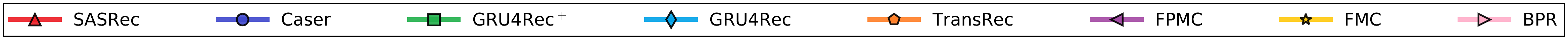}
\end{subfigure}
\begin{subfigure}[b]{0.245\textwidth}
\includegraphics[width=\linewidth]{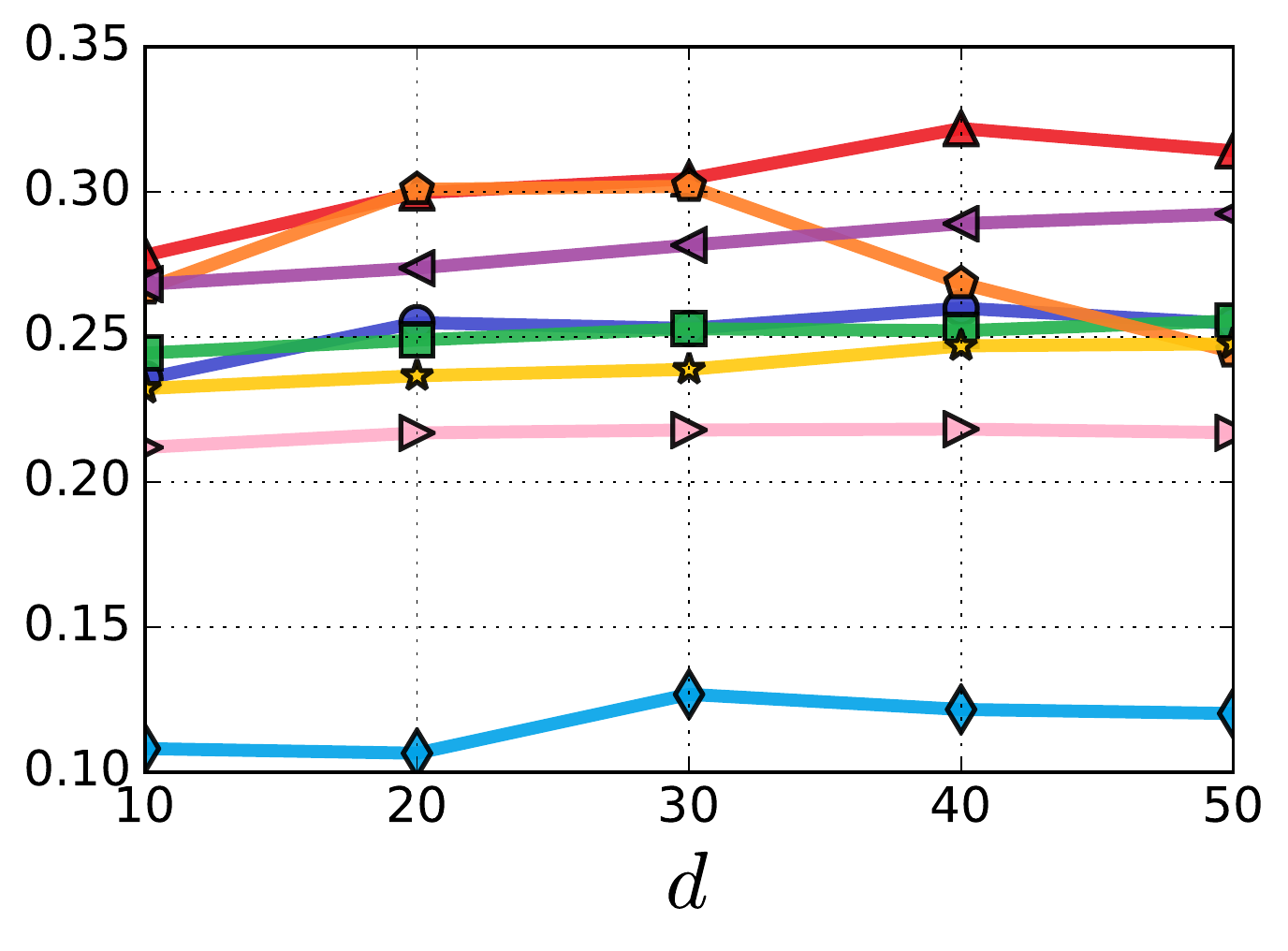}
\subcaption{\emph{Beauty}}
\end{subfigure}
\begin{subfigure}[b]{0.245\textwidth}
\includegraphics[width=\linewidth]{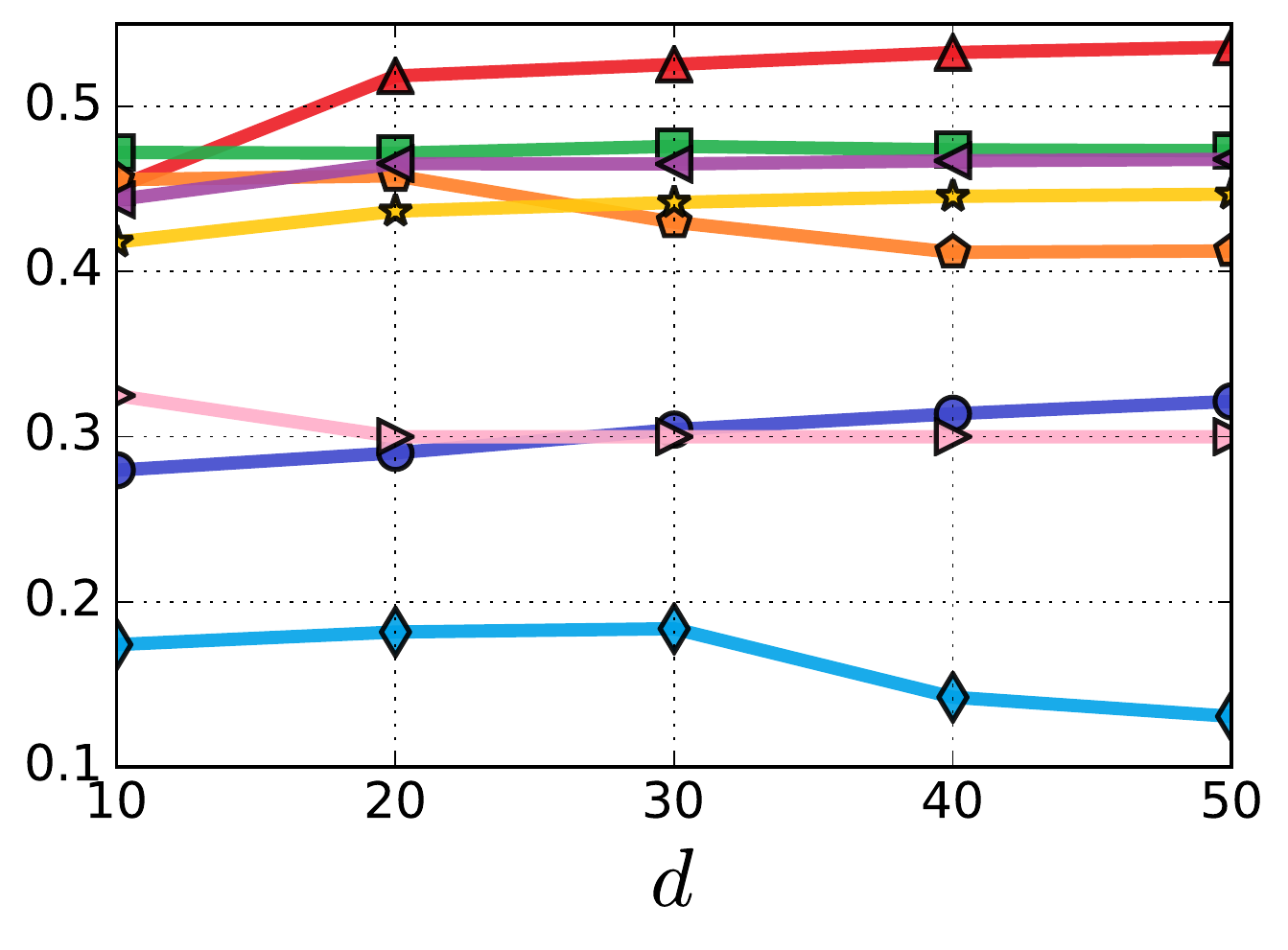}
\subcaption{\emph{Games}}
\end{subfigure}
\begin{subfigure}[b]{0.245\textwidth}
\includegraphics[width=\linewidth]{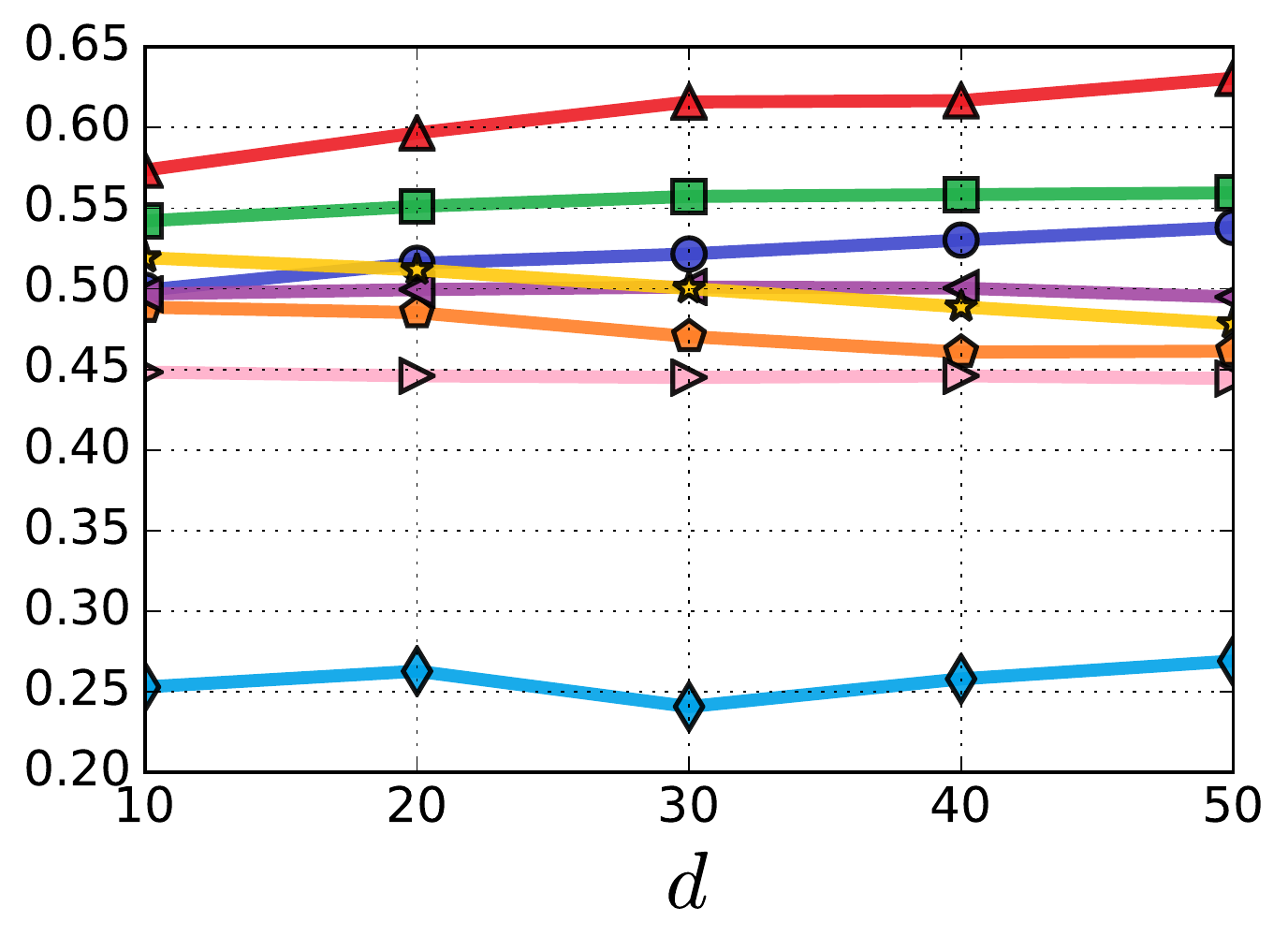}
\subcaption{\emph{Steam}}
\end{subfigure}
\begin{subfigure}[b]{0.245\textwidth}
\includegraphics[width=\linewidth]{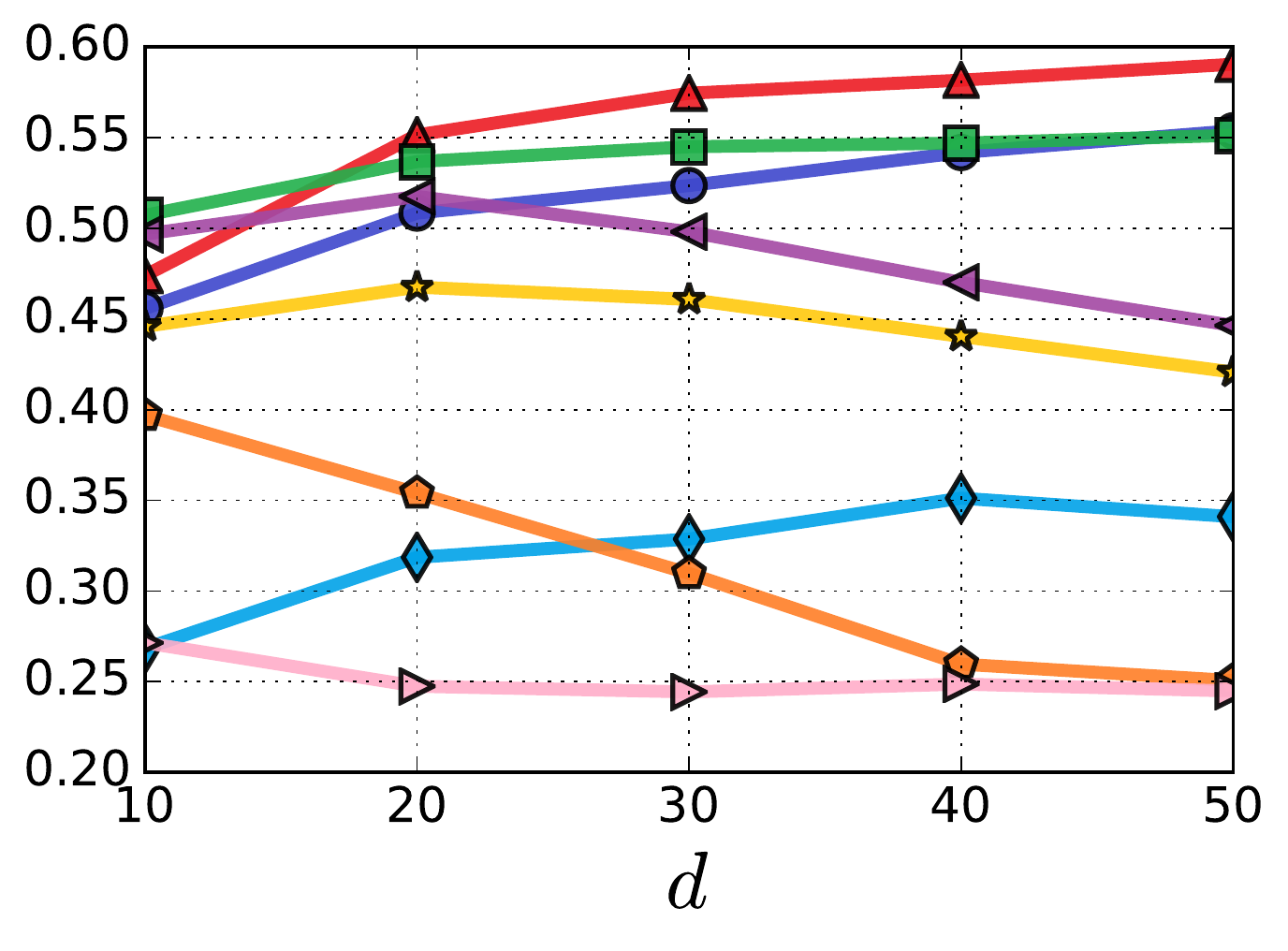}
\subcaption{\emph{ML-1M}}
\end{subfigure}
\caption{Effect of the latent dimensionality $d$ on ranking performance (NDCG@10).
}
\label{fig:K}
\end{figure*}

In Figure \ref{fig:K} we also
analyze the effect of 
a
key hyper-parameter, the latent dimensionality $d$, by showing NDCG@10 of all methods with $d$ varying from 10 to 50. We see that our model typically benefits from larger numbers of latent dimensions. For all datasets, our model achieves satisfactory performance with $d\geq40$.

\subsection{Ablation Study}

Since there are many components in our architecture, we analyze their impacts via an ablation study (\textbf{RQ2}). Table \ref{tb:ablation} shows the performance of our default method and its \textbf{8} variants on all four dataset (with $d=50$). We introduce the variants and analyze their effect respectively:

\begin{table}[h]
\centering
\caption{Ablation analysis (NDCG@10) on four datasets. Performance better than the default version is boldfaced. `$\downarrow$' indicates a severe performance drop (more than 10\%).}
\begin{tabular}{lllll}
\toprule
Architecture				& \emph{Beauty}		&	\emph{Games}	&	\emph{Steam}	&	\emph{ML-1M}\\ \midrule                               
(0) Default  				&	0.3142				&	0.5360					&	0.6306				&	0.5905 \\
(1) Remove PE        	 	&	\textbf{0.3183}		&	0.5301					&	0.6036				&	0.5772 \\
(2) Unshared IE				&	0.2437$\downarrow$	&	0.4266$\downarrow$		&	0.4472$\downarrow$	&	0.4557$\downarrow$ \\
(3) Remove RC        	 	&	0.2591$\downarrow$	&	0.4303$\downarrow$		&	0.5693	&	0.5535 \\
(4) Remove Dropout         	&	0.2436$\downarrow$	&	0.4375$\downarrow$		&	0.5959	&	0.5801 \\
(5) 0 Block ($b$=0)			&	0.2620$\downarrow$  &	0.4745$\downarrow$		&	0.5588$\downarrow$	&	0.4830$\downarrow$ \\
(6) 1 Block ($b$=1)			&	0.3066 				&	\textbf{0.5408}			&	0.6202	&	0.5653 \\
(7) 3 Blocks ($b$=3)			&	0.3078			 	&	0.5312					&	0.6275	&	\textbf{0.5931} \\
(8) Multi-Head 			  	&	0.3080				&	0.5311					&	0.6272	&	0.5885 \\
\bottomrule
\end{tabular}
\label{tb:ablation}
\end{table}

\begin{itemize}
\item (1) \emph{Remove PE (Positional Embedding)}: Without the positional embedding $\P$, the attention weight on each item 
depends only on item embeddings.
That is to say, the model makes recommendations 
based on users' past actions,
but
their order doesn't matter.
This
variant might be suitable for sparse datasets, where user sequences are typically short. 
This variant performs better then the default model on the sparsest dataset (\emph{Beauty}), but worse on other denser datasets.
\item (2) \emph{Unshared IE (Item Embedding)}: We 
find
that using two item embeddings consistently impairs the performance,
presumably due to overfitting.

\item (3) \emph{Remove RC (Residual Connections)}: 
Without residual connections, we find that performance is significantly worse. 
Presumably this is because
information in lower layers (e.g.~the last item's embedding and the output of the first block) can not be easily propagated to the final layer, and this information is highly useful for making recommendations, especially on sparse datasets.

\item (4) \emph{Remove Dropout}: 
Our results show that
dropout can effectively regularize the model to achieve better test performance, especially on sparse datasets. The results also imply the overfitting problem is less severe on dense datasets.

\item (5)-(7) \emph{Number of blocks}:
Not surprisingly, results are inferior
with
zero blocks, since the model would only depend on the last 
item. The variant with one block performs reasonably well, though using two blocks (the default model) still boosts performance especially on dense datasets, meaning that the hierarchical self-attention structure is helpful to learn more complex item transitions. Using three blocks achieves similar performance to the default model.

\item (8) \emph{Multi-head}: 
The authors
of Transformer~\cite{transform} found that it is useful to use `multi-head' attention, which 
applies
attention in $h$ subspaces (each a $d/h$-dimensional space). However,
performance with two heads is consistently and slightly worse than single-head attention in our case. 
This might owe
to the small $d$ in our problem ($d=512$ in Transformer), which is not suitable for decomposition into smaller subspaces.


\end{itemize}

\subsection{Training Efficiency \& Scalability}

We evaluate two aspects of the training efficiency (\textbf{RQ3}) of our model: 
Training speed (time taken for one epoch of training) and convergence time (time taken to achieve satisfactory performance). 
We also 
examine
the scalability of our model in terms of the maximum length $n$. All experiments are conducted with a single GTX-1080 Ti GPU.

\begin{figure}[h]
\centering
\includegraphics[width=.40\textwidth]{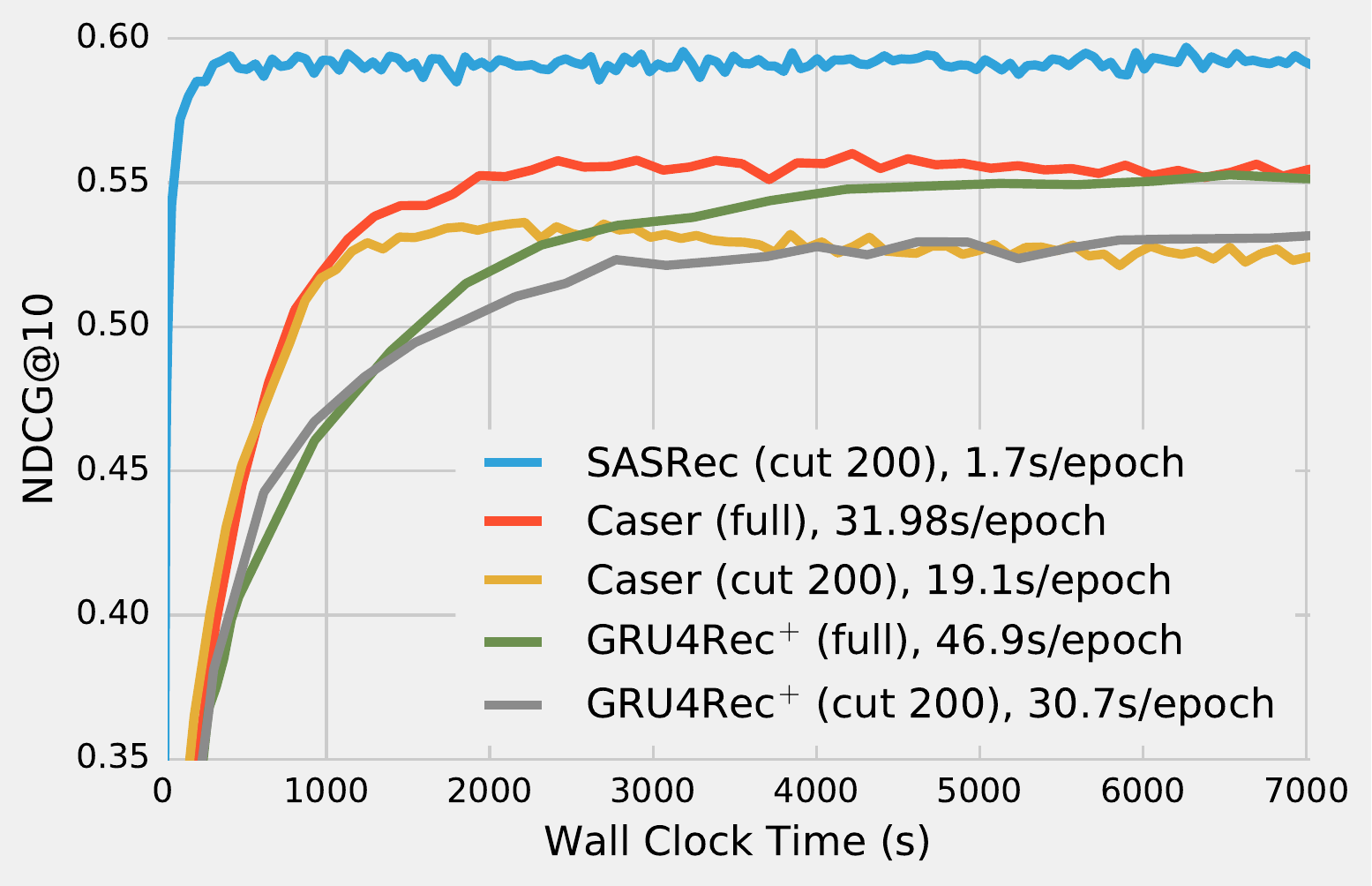}
\caption{Training efficiency on \emph{ML-1M}. 
SASRec is an order of magnitude faster than CNN/RNN-based recommendation methods in terms of training time per epoch and in total.}
\label{fig:speed}
\end{figure}

\xhdr{Training Efficiency} Figure \ref{fig:speed} demonstrates the efficiency of deep learning based methods with GPU acceleration. GRU4Rec is omitted due to its 
inferior
performance. For fair comparison, there are two training options for Caser and GRU4Rec$^{\text{+}}$: using 
complete
training data or just the most recent 200 actions (as in SASRec). 
For computing speed, SASRec only spends 1.7 seconds on model 
updates
for one epoch, which is over 11 times faster than Caser (19.1s/epoch) and 18 times faster than GRU4Rec$^{\text{+}}$ (30.7s/epoch). 
We also see that
SASRec 
converges to optimal 
performance within around 350 seconds on \emph{ML-1M} while other models require much 
longer.
We also find that
using full data 
leads to better
performance for Caser and GRU4Rec$^{\text{+}}$.

\begin{figure*}
\centering
\begin{subfigure}[b]{0.245\textwidth}
\includegraphics[width=\linewidth]{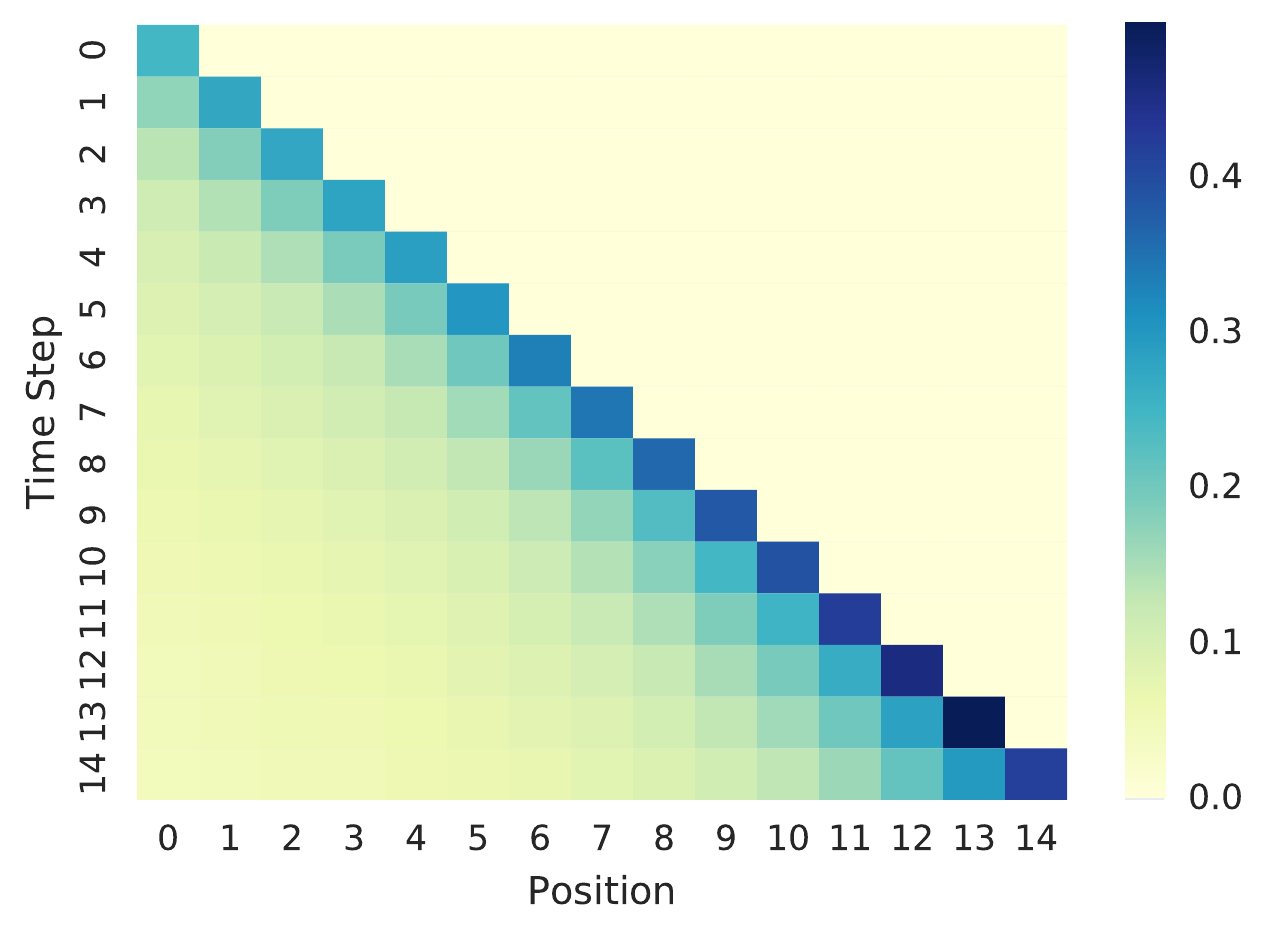}
\subcaption{\emph{Beauty}, Layer 1}
\end{subfigure}
\begin{subfigure}[b]{0.245\textwidth}
\includegraphics[width=\linewidth]{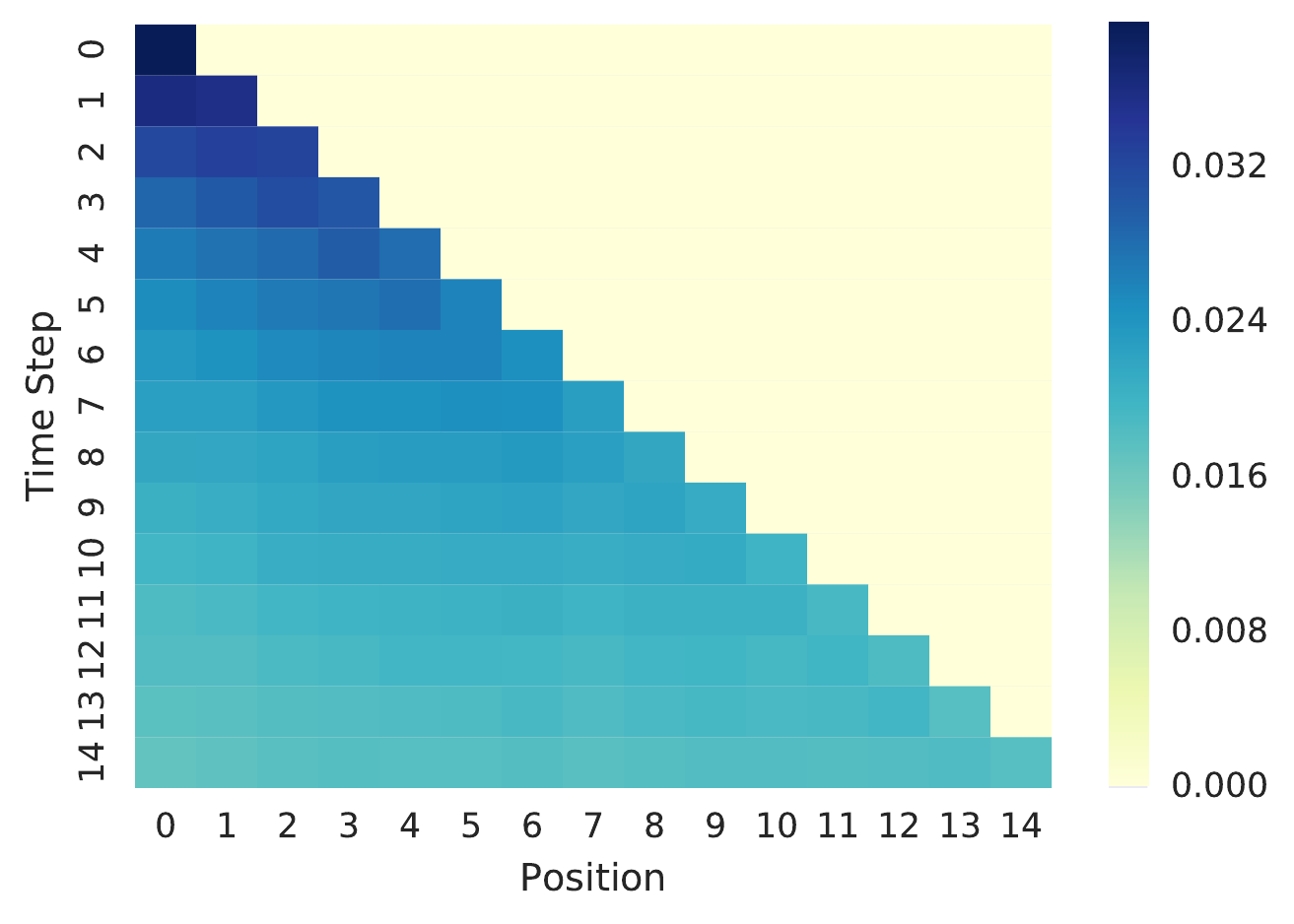}
\subcaption{\emph{ML-1M}, Layer 1, w/o PE}
\end{subfigure}
\begin{subfigure}[b]{0.245\textwidth}
\includegraphics[width=\linewidth]{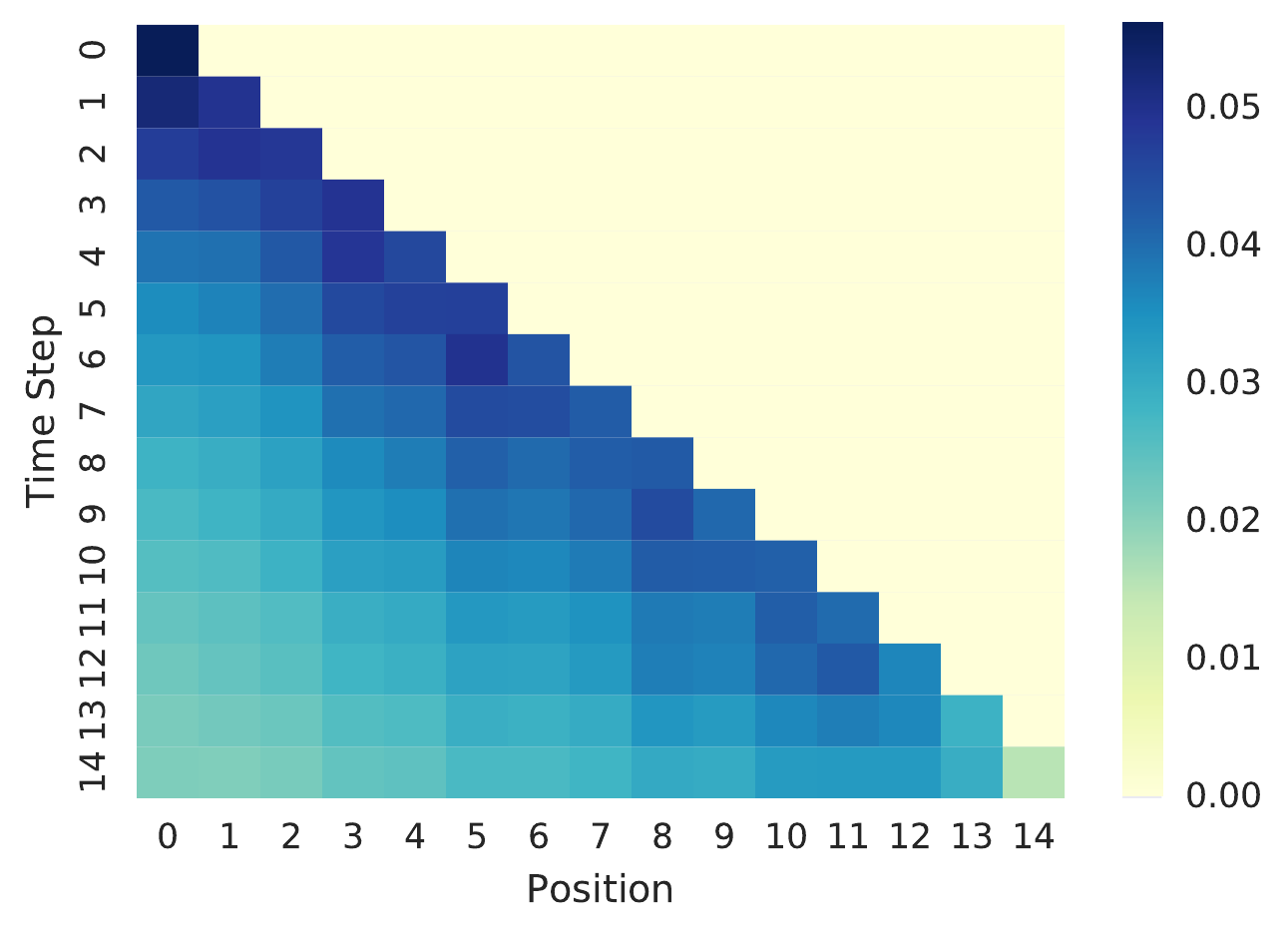}
\subcaption{\emph{ML-1M}, Layer 1}
\end{subfigure}
\begin{subfigure}[b]{0.245\textwidth}
\includegraphics[width=\linewidth]{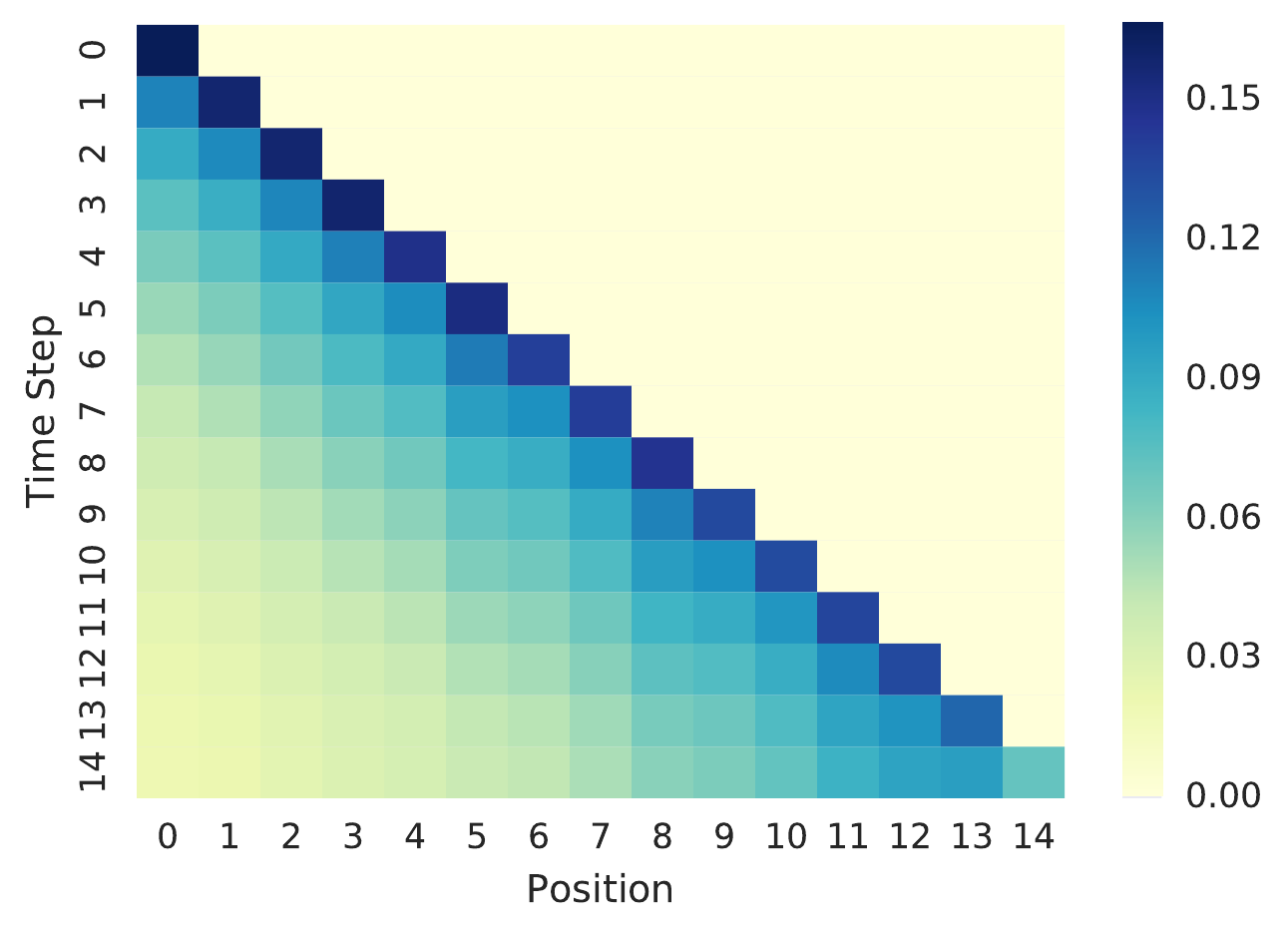}
\subcaption{\emph{ML-1M}, Layer 2}
\end{subfigure}
\caption{Visualizations of average attention weights on positions at different time steps. For comparison, the heatmap of a first-order Markov chain based model would be a diagonal matrix.}
\label{fig:vis_pos}
\end{figure*}

\xhdr{Scalability} As with standard MF methods, SASRec scales linearly with the total number of users, items and actions. A potential scalability concern is 
the maximum length $n$, however
the computation can be effectively parallelized
with GPUs. 
Here
we measure the training time and performance of SASRec with different $n$, empirically study its scalability, and analyze whether it can 
handle sequential recommendation in most cases. Table \ref{tab:scale} shows the performance and efficiency of SASRec with various sequence lengths. 
Performance is
better with larger $n$,
up to around $n=500$ at which point
performance saturates (possibly because 99.8\% of actions have been covered). However, even with $n=600$, the model can be trained in 2,000 seconds, which is still faster than Caser and GRU4Rec$^{\text{+}}$. Hence, our model can easily scale to user sequences 
up to a few hundred actions, which is suitable for typical review and purchase datasets.
We plan to investigate approaches (discussed in Section \ref{sec:complexity}) for handling very long sequences
in the future.


\begin{table}[h]
\centering
\caption{Scalability: performance and training time with different maximum length $n$ on \emph{ML-1M}.}
\setlength{\tabcolsep}{3pt}
\begin{tabular}{lccccccccc}
\toprule
$n$							&	10	&	50	&	100	&	200	&	300	&	400	&	500	&	600\\ \midrule                               
Time(s)  					&	75 	&	101 	&	157 	&	341 	&	613 	&	965 	&	1406 	&	1895 	&	\\
NDCG@10 					&	0.480 	&	0.557 	&	0.571 	&	0.587 	&	0.593 	&	0.594 	&	0.596 	&	0.595  \\                      
\bottomrule
\end{tabular}
\label{tab:scale}
\end{table}

\subsection{Visualizing Attention Weights}
\label{sec:vis}

Recall that at time step $t$, the self-attention mechanism in our model adaptively assigns weights on the first $t$ items depending on their position embeddings and item embeddings. To answer 
\textbf{RQ4}, we examine all 
training sequences and seek to reveal meaningful patterns by showing the average attention weights on positions as well as items.

\xhdr{Attention on Positions}

Figure \ref{fig:vis_pos} shows four heatmaps of average attention weights on the last 15 positions at the last 15 time steps. Note that when we calculate 
the 
average weight, the denominator is 
the
number of \emph{valid} weights, 
so as to
avoid the influence of padding items in short sequences.

We consider a few comparisons among the heatmaps:
\begin{itemize}
\item \emph{(a) vs.~(c)}: This comparison indicates that the model tends to attend on more recent items on the sparse dataset \emph{Beauty}, 
and less recent items for 
the dense dataset \emph{ML-1M}.
This is the key factor 
that allows our model
to adaptively handle both sparse and dense datasets, 
whereas existing methods tend to focus on one end of the spectrum.

\item \emph{(b) vs.~(c)}: This comparison shows the effect of using positional embeddings (PE). Without 
them
attention weights 
are essentially
uniformly distributed over previous items, while the default model (c) is more sensitive in position 
as it is
inclined to attend on recent items. 

\item \emph{(c) vs.~(d)}: Since our model is hierarchical, this shows how 
attention 
varies
across different blocks. Apparently, attention in high layers tends to focus on more recent positions. 
Presumably 
this is
because the first self-attention block already considers all previous items, and the second block does not need to consider far away positions. 
\end{itemize}

Overall, the visualizations show that the behavior of our self-attention mechanism is \emph{adaptive}, \emph{position-aware}, and \emph{hierarchical}. 

\xhdr{Attention Between Items}

\begin{figure}[h]
\centering
\includegraphics[width=.45\textwidth]{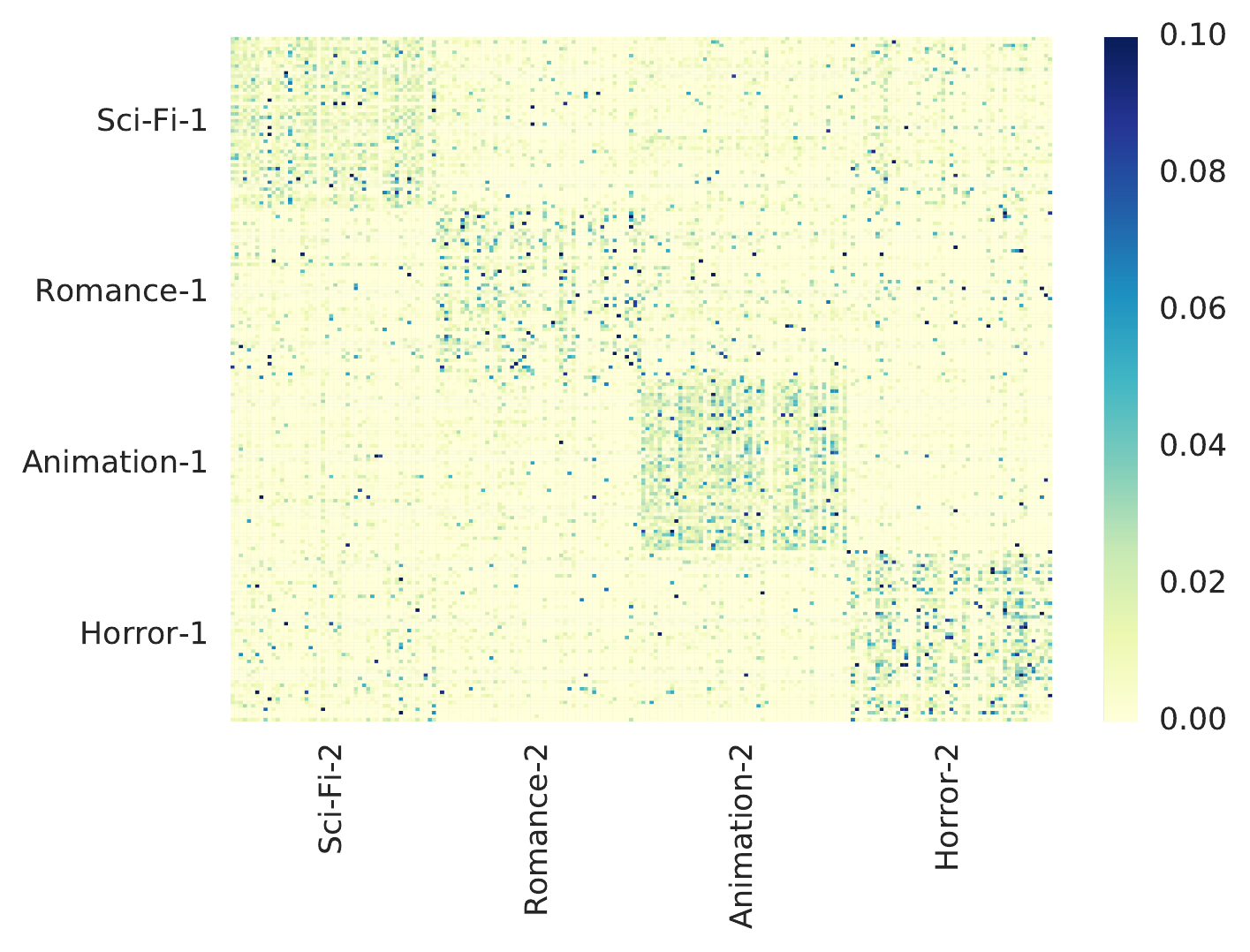}
\caption{Visualization of average attention between movies from four categories. This shows our model can uncover items' attributes, and assigns larger weights between similar items.}
\label{fig:vis_item}
\end{figure}

Showing attention weights between a few cherry-picked items might not be statistically meaningful. To perform a broader comparison, using \emph{MovieLens-1M}, where each movie has several categories, we randomly select two disjoint sets where each set contains 200 movies 
from 4 categories: \emph{Science Fiction (Sci-Fi)}, \emph{Romance}, \emph{Animation}, and \emph{Horror}. The first set is used for the query and the second set 
as the key. Figure \ref{fig:vis_item} shows a heatmap of average attention weights between the two sets. We can see the heatmap is approximately a block diagonal matrix, meaning that the attention mechanism can identify similar items (e.g.~items sharing a common category) and tends to assign larger weights between them (without being aware of categories in advance).

\section{Conclusion}

In this work, we proposed a novel self-attention based sequential model \emph{SASRec} for next item recommendation. SASRec models the entire user sequence (without any recurrent or convolutional operations), and adaptively considers consumed items for prediction. Extensive empirical results on both sparse and dense datasets show that our model outperforms state-of-the-art baselines, and is an order of magnitude faster than CNN/RNN based approaches.
%
In the future, we plan to extend the model by incorporating rich context information (e.g.~dwell time, action types, locations, devices, etc.), and to investigate approaches to handle very long sequences (e.g.~clicks).

\footnotesize
\bibliographystyle{IEEEtran}
\bibliography{sigproc} 

\end{document}